# Mixed Nonlinear Response and Transition of Nonlinearity in a Piezoelectric Membrane

Nishta Arora[#], Priyanka Singh[#], Randhir Kumar, Rudra Pratap, Akshay Naik*

*Centre for Nano Science and Engineering, Indian Institute of Science, Bengaluru, 560012, India*

*Email: anaik@iisc.ac.in

## Abstract

**Nonlinearities play a critical role in the dynamics of mechanical resonators, enhancing sensitivity and enabling signal manipulation. Understanding the parameters affecting nonlinearities is crucial for developing strategies to counteract their effects or manipulate them for improved device performance. This study investigates the impact of fabrication-induced curvature on the dynamics of zinc oxide-based piezoelectric micromachined ultrasonic transducers (PMUTs). Our experiments reveal that these devices exhibit hardening, softening, and mixed nonlinear responses, with varying initial static displacements. Notably, PMUTs with almost flat initial static displacement exhibit hardening nonlinearity, while those with a curved initial static displacement show softening nonlinearity. An exotic mixed nonlinear response is observed for intermediate static displacement. We attribute the observed nonlinear response to the interplay between static displacement-induced quadratic nonlinearity and midplane stretching-induced cubic nonlinearity. We provide a theoretical formulation for the dynamics of the devices, which explains the experimental results and highlights the nonlinear responses and their dependence on the initial static displacement. Our findings underscore the significance of nonlinearities in the dynamics of mechanical resonators and suggest ways to optimize device performance.**

Micro-electromechanical system (MEMS) based devices have been extensively studied and used in applications such as inertial navigation[1], digital light manipulation (e.g. DMDs), radio frequency (RF) communication[2], biosensors[3] and several types of transducers[4]. MEMS based ultrasonic transducers are of immense interest for their application in proximity sensing[5], gesture recognition[6], medical imaging[7], photoacoustics[8] and data-over-sound communication[9].

These transducers are preferred over conventional bulk ultrasonic transducers due to low power consumption[10], low cost, and good acoustic impedance matching[10]. These transducers are generally of two types: capacitive micromachined ultrasonic transducer (CMUTs)[11,12] and piezoelectric ultrasonic transducer (PMUTs)[13]. CMUTs have several drawbacks which limit their applications[14]. These include a large bias voltage for operation and an external circuit for the capacitance measurement. PMUTs overcome these disadvantages and are used in many applications, including 2D PMUT array for intravascular ultrasound (IVUS)[15], ultrasonic fingerprint sensor based on an array of PMUT[16,17], water propagation in micropumps[18], sonography[19], air propagation in range finder[20,21], gesture recognition[22], mass sensors[23] and for fluid density sensing using both static and dynamic modes[24]. In all these applications, PMUTs are operated in the linear response region. However, with miniaturization and efficient actuation schemes, these devices can easily achieve large vibration amplitudes and exhibit nonlinear responses. To exploit their potential in various applications, it is essential to understand the nonlinear response and its controllability.

Nonlinearity in mechanical structures has been exploited to enhance the device performance. Choi et al.[25] have shown that introducing nonlinearity through structural modification can improve the frequency sensitivity of a magnetic sensor. Kacem et al.[26] have demonstrated that improved dynamic range by operating in the nonlinear region. Samanta et al.[27] have also manipulated nonlinearities in ultrathin membranes to increase the dynamic range. Enhanced dynamic range has implications for improving sub-single-atom resolution in NEMS mass spectrometry[28]. Sina et al.[30] have found that the sensitivity of a curved PMUT structure is better than that of a planar configuration. By taking advantage of the tunability of nonlinearity, it is possible to enhance device sensitivity, and other performance metrics. Understanding and controlling nonlinearity is crucial for achieving optimal performance from PMUTs.

Several factors, such as fabrication process[31], material nonlinearity[32], large amplitude vibration[27], and damping environment[33,34] can introduce nonlinearity in MEMS devices. The microfabrication process for PMUTs leads to several imperfections due to stress gradients in the film stack of the structure. These stress gradients can lead to curvature in these devices. Amabili et.al.[35] have studied the effect of curvature on the nonlinear response of shallow shells with rectangular base subjected to harmonic excitation. Capacitive actuation of beams, membranes, and CMUTs have been studied to demonstrate tuning of nonlinearity with applied DC bias[27,36,37]. However, similar studies with PMUTs have not been explored. PMUTs require efficient actuation with high-quality oriented piezo material to achieve large vibrational amplitudes. Using a microwave-assisted technique[38], we can grow highly oriented ZnO as the device layer for our PMUT arrays. The resulting PMUT arrays have large amplitudes and exhibit nonlinear responses, providing an opportunity for in-depth analysis of their behavior.

In this work, we investigate the effect of the initial static displacement in the form of the initial curvature of the circular PMUT membrane on its nonlinear dynamic response. The predominant contribution to this initial static displacement is due to the fabrication induced stress gradients. In some cases, this static displacement can also be modified by annealing the device. We observe hardening, softening and mixed nonlinear responses in different devices with varying initial static displacement. The nonlinear response observed experimentally is a function of each device's initial static displacement. We provide a theoretical model to understand and validate the experimentally observed effect of initial static displacement on the dynamics of these devices.

Here, we use zinc-oxide based MEMS ultrasonic transducers (PMUTs). These devices with circular diaphragm are fabricated on SOI using micromachining techniques[39]. The simplified fabrication process flow for ZnO PMUTs is shown in Figure 1(a). The SOI chip, with ZnO layer on top, is etched from the bottom side to obtain a diaphragm of different diameters. The top electrode partially covers (66%) this diaphragm. This coverage is ideal for exciting the first axisymmetric vibrational mode using a single electrode[40]. Further details of the device fabrication are discussed in the supplementary material (section A). The optical image of a 6x6 cm$^2$ chip with several devices on it is shown in fig. 1(b). Each chip has devices of four different diameters: 500μm, 1000μm, 2000μm and 3000μm with a common bottom electrode. Figure 1(c) shows the top view of a single PMUT device. Figure 1(d) shows the simplified schematic of the fabricated PMUT with a flat and curved configuration. The deposition of multiple thin films during fabrication process leads to a stress gradient in the film stack. The variability in diaphragm's static displacement is due to this stress. Here we study the effect of this static displacement (quantified by $w_{max}$) on the dynamic response of the device.

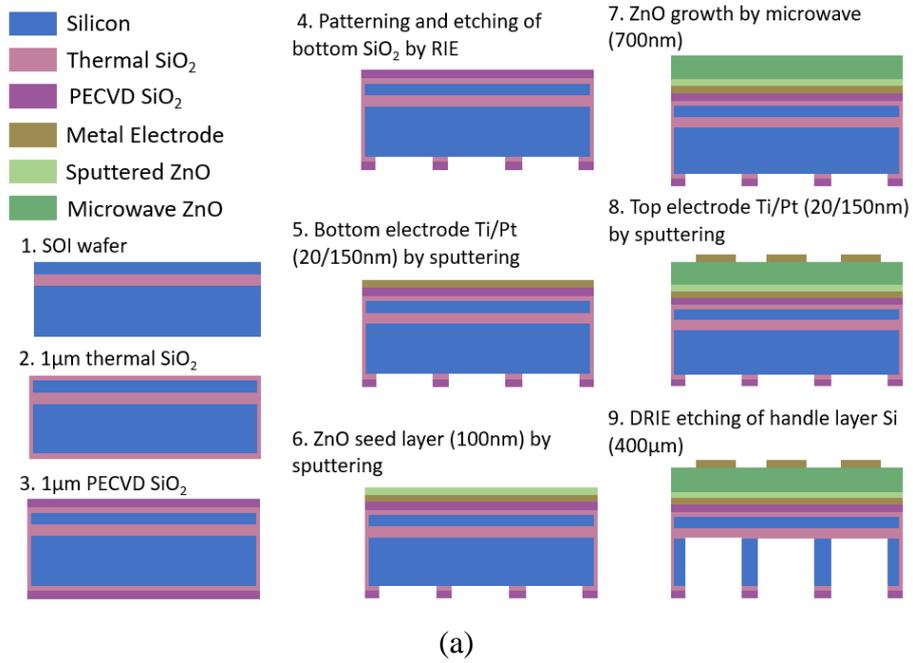

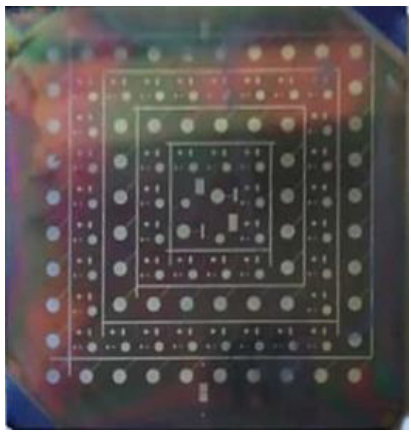

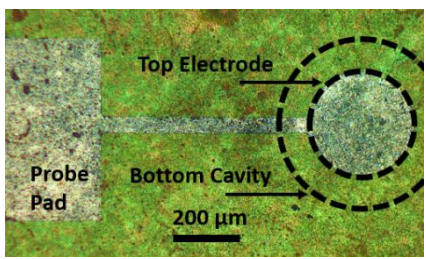

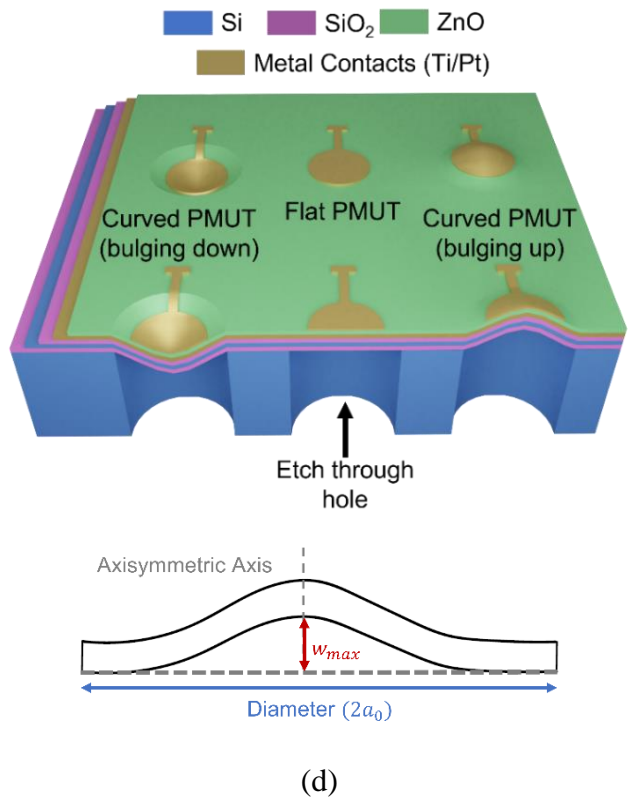

**Figure 1:** (a) Cross-section view of the fabrication process flow for ZnO PMUTs. (b) Optical image of arrays of PMUT devices with varying diameters (c) Single device with diaphragm diameter of 500μm and top electrode partially covering the device. (d) Schematic shows the constituent layers of ZnO PMUT fabricated on SOI wafers with different possible configurations. Below is zoomed in view of PMUT bulging upwards with initial static displacement $w_{max}$.

The static displacement of PMUTs is measured using an optical profilometer (TalySurf CCI). Figure 2(a) shows the 3D topography of device D1. The dark blue region corresponds to the reflection in the optical profilometer observed from the bottom metal electrode. The ZnO layer on top of this bottom electrode is transparent at the wavelengths used in the optical profilometer. The light blue region in fig. 2(a) is from the diaphragm region containing the top electrode. The 2D profile of the same device measured along the device diameter is shown in the supplementary information (fig. S1). D1 is almost flat, with a step height of around $\approx 1 \ \mu m$. This step height corresponds to the thickness of 800 nm ZnO and 170 nm of top metal electrode. Figure 2(b) shows the 3D topography of another PMUT (device D2). The diaphragm of this device has an initial static displacement arising from fabrication nonidealities. The 2D profile along the diameter of the PMUT is shown in the supplementary information (fig.S2). We have used the 2D profile to calculate the initial static displacement at the centre of the diaphragm ($w_{max}$). This static displacement of the diaphragm has been corrected for the additional thickness due to ZnO and top metal layer. For device D1, $w_{max}$ is calculated to be nominally zero, while for device D2, it is estimated to be 1.02 μm (0.20% of diameter). Details of profilometer measurements on other devices are given in the supplementary information.

We have characterized the dynamic response of multiple ZnO diaphragms with 500 μm diameter in air and at room temperature. The dynamic response to the piezoelectric actuation is measured using Micro Scanning Laser Vibrometer (model: Polytec MSV 500). A simplified measurement schematic is shown in fig. 2(c). The voltage signal generated by the internal signal generator of the LDV is applied between the top and the bottom electrode to excite the out-of-plane motion of the diaphragm. A periodic chirp input signal with uniform energy distribution from DC to 2 MHz is used to excite the PMUT. This enables us to identify the resonant frequencies and the corresponding modal shapes of the flexural modes. Figure 2(d) shows the response of the device measured using the LDV. All subsequent measurements in this work have been performed on the fundamental mode of the device.

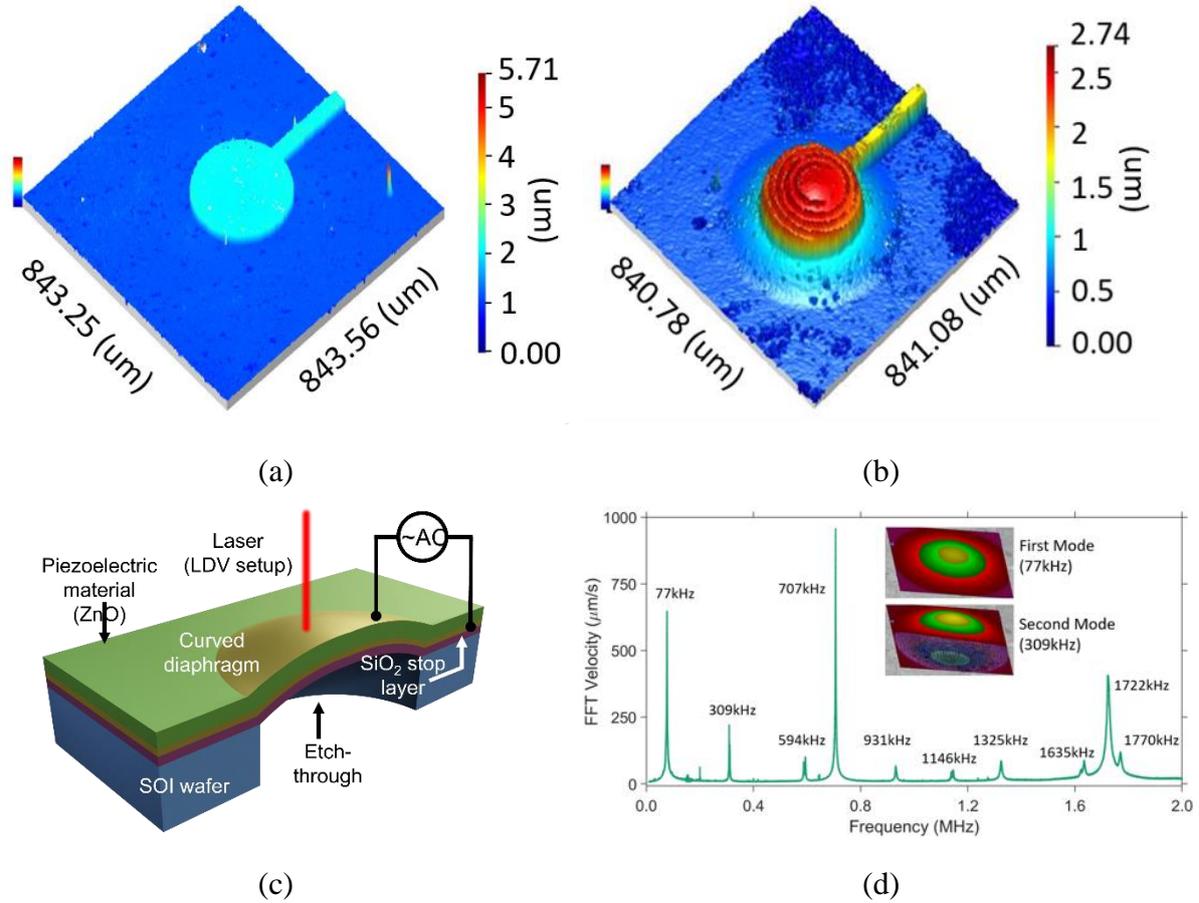

**Figure 2:** 3D topography of (a) flat device (D1) and (b) curved device (D2). (c) Cross-sectional schematic and measurement setup for ZnO PMUT. A voltage signal is applied between the electrodes to actuate the device. (d) The electromechanical response, of the 500μm diameter PMUT, to a periodic chirp signal ($V_{ac} = 2V$). Inset shows the mode shapes of the first two out of plane modes.

To study the frequency response, a sinusoidal voltage ($V_{ac}$) is applied to the electrodes and the actuation frequency is swept around the resonant frequency of the first mode. Figure 3(a) shows the experimentally observed response of the device D1. At low actuation voltages, the diaphragm's response is linear and at higher voltages it exhibits a hardening nonlinear response due to the mid-plane stretching. Similar measurements were performed on device D2, which had an initial static displacement of ~1μm (0.20% of diameter). Figure 3(b) shows the experimental frequency response curve with increasing actuation drive ($V_{ac}$). We observe a linear response at lower actuation drive and softening nonlinear response on increasing the actuation voltage. This shows that the effect of quadratic nonlinearity is more than the cubic nonlinearity, which leads to negative Duffing constant. Since the devices are nominally similar except for the initial static displacement, the measurements allude to the effect of the initial curvature on the nonlinearities observed.

To explain the observed experimental response, we have developed a theoretical model that provides insights into the underlying physics of the system. Since the silicon and ZnO layers are thicker than the other layers and are likely to play a significant role in the system's dynamics, we have considered only these two layers in our model (see supplementary information section C). By assuming large displacement amplitude (that is, incorporating the in-plane forces), imposing the condition of axial symmetry[41,42], and assuming the initial static displacement in the system, we obtain the following equation of motion for PMUT[43]:

$$\frac{D}{a_0^4}\nabla^4 w_0(r) + \frac{D}{a_0^4}\nabla^4 w_1(r) + c\frac{\partial}{\partial t}w_1(r,t) + \rho_s \frac{\partial^2}{\partial t^2}w_1(r,t)$$
$$-\frac{N}{a_0^4}\left(\int_0^{a_0} r\left(\left(\frac{\partial w_1(r,t)}{\partial r}\right)^2 + 2\frac{\partial w_0(r)}{\partial r}\frac{\partial w_1(r)}{\partial r}\right)dr\right)\left(\frac{\partial^2(w_0(r)+w_1(r,t))}{\partial r^2} + \frac{1}{r}\frac{\partial^2(w_0(r)+w_1(r,t))}{\partial r^2}\right) = F$$

(1)

where, $w(r,t) = w_0(r) + w_1(r,t)$ is the total displacement of the device, $w_0(r)$ represents the initial static displacement and $w_1(r,t)$ represent the dynamic part of the vibrations.

The solution of eqn. (1) can be sought in terms of the linear superposition of normal vibration modes:

$$w_1(r,t) = \sum u_{mn}(t)\psi_{mn}(r) \quad (2)$$

where $u_{mn}(t)$ is the coordinate factor and $\psi_{mn}(r)$ is the mode shape factor. The mode shape factor can be written as

$$\psi_{mn}(r) = J_n(\alpha_{mn}r) - \frac{J_n(\alpha_{mn})}{I_n(\alpha_{mn})}I_n(\alpha_{mn}r) \quad (3)$$

Here $\alpha_{mn}$ is determined by solving the transcendental equation $\alpha_{mn}\frac{J_{n+1}(\alpha_{mn})}{J_n(\alpha_{mn})} + \alpha\frac{I_{n+1}(\alpha_{mn})}{I_n(\alpha_{mn})} = 0$

, where $J_n$ and $I_n$ are Bessel functions of the second and first kind, respectively.

Using this and further simplifications (see supplementary information section C), we obtain the equation of motion of the PMUT as :

$$\ddot{u} + 2\xi\dot{u} + u + b_2 u^2 + b_3 u^3 = F \quad (4)$$

We assume that the initial static displacement is axisymmetric with the condition that the static displacement should be zero at the circumference, i.e., at $r = a_0$, $w = 0$. At the centre of the diaphragm, the static displacement is assumed to be maximum, i.e., at $r = 0$, $w = w_{max}$ and the equation of the curve is given by

$$w_0 = w_{max}\left[1-\left(\frac{r}{a_0}\right)^2 + k\left(\frac{r}{a_0}\right)\ln\left(\frac{r}{a_0}\right)\right] \tag{5}$$

Where, $k$ represents the steepness of the curve. The value of $k$ is obtained by fitting the experimentally measured curvature of each device.

Using the known experimental parameters, eqn. (4) can be used to model the dynamic response of ZnO PMUTs (see supplementary information for further details). In this equation, the quadratic and cubic term represents the strain nonlinearity due to initial static displacement and geometric nonlinearity due to midplane stretching of the diaphragm, respectively. The solution of Eqn. (4) is obtained numerically in MATLAB by using the ODE Solver ODE45.

The results of the numerical model for device D1 and D2 are shown in fig. 3(c) and 3(d) respectively. The numerical simulations closely align with the experimentally observed nonlinear response of the two devices. The effect of initial static displacement is to introduce quadratic nonlinearity. The model predicts that, for small initial static displacement, the cubic nonlinearity due to midplane stretching is the dominant nonlinear effect. As the initial static displacement increases, the effect of quadratic nonlinearity on the dynamical response becomes dominant and the device exhibits softening nonlinearity.

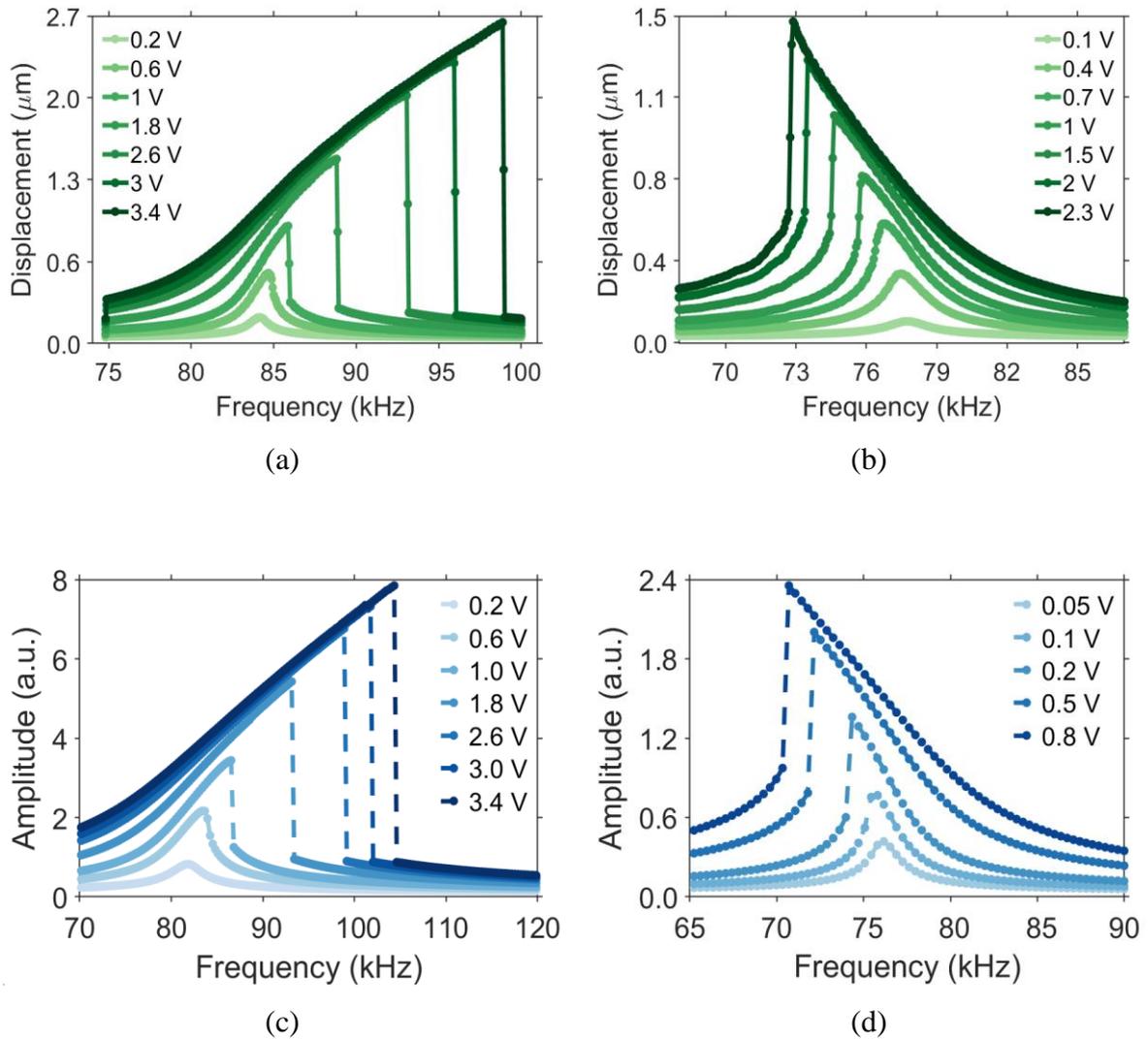

**Figure 3**: (a) Experimental frequency response of the fundamental mode with increasing AC drive of (a) a flat PMUT (D1) diaphragm, showing hardening nonlinearity and (b) a curved PMUT (D2) diaphragm with initial static displacement, depicting softening nonlinearity. (c) and (d) show the simulated frequency response with increasing AC drive obtained from the theoretical model for device D1 and D2 respectively.

To further probe the effect of initial static displacement on the nonlinear response and understand the complete device dynamics, we have performed measurements on device D2 at actuation voltages higher than 2.3V. At these higher actuation voltages, we observe both softening and hardening nonlinearity with bifurcation during the forward frequency sweep (fig. 4(a)). Similar response, but with different frequency jump points, are also observed in the reverse frequency sweep. In device D2, multiple frequency jumps during forward and reverse sweeps are due to the interplay of both quadratic and cubic nonlinearity. The simulated mixed nonlinear response at high actuation voltage for device D2 is shown in fig. 4(b). To estimate the initial static displacement required to observe mixed nonlinear response, we simulate the response for different PMUT radii. The static displacement required for observation of mixed nonlinear response increases on increasing the radius of PMUT (refer supplementary information fig. S6 for details). In device D1, which has a very small initial static displacement, contribution from quadratic nonlinearity is very small and the dynamic response exhibits only hardening nonlinearity.

Additionally, to confirm the observation of transition of nonlinearity with initial static displacement, we study the dynamic response of another device (D3) before and after annealing. This device has an initial static displacement of $w_{max}$ of $0.55 \mu m$ (0.11% of the diameter) due to fabrication induced stresses. The PMUT device stack comprises of different layers having a varying coefficient of thermal expansion (CTE). Annealing these devices can leads to uneven film stresses and affect the overall diaphragm stiffness and resonant frequency. Since the device is clamped at the edges, it also leads to a significant change in the static displacement of the device. Figure 5(a) shows the measured 2D profile of the PMUT (device D3) before (green curve) and after (orange curve) annealing. The device was annealed in an oxygen environment in an electrical heating furnace at $800^0$C for three hours. The PMUT diaphragm shows a clear transition from an almost flat (static displacement ~ 0.55μm (0.11% of diameter)) to a curved membrane structure with a significantly greater initial static displacement of 2.23μm (0.45 % of diameter).

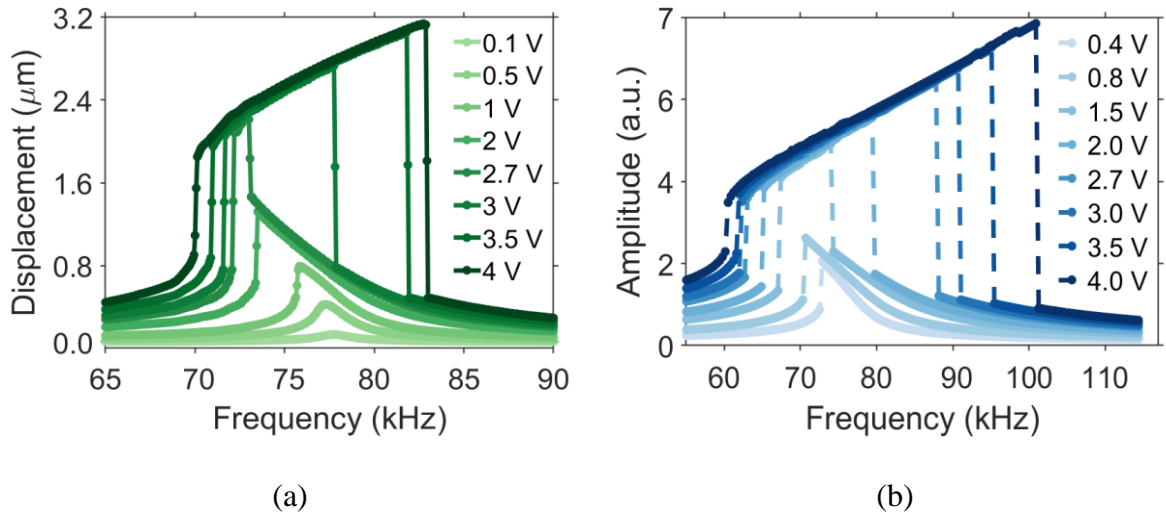

(a)          (b)

**Figure 4:** (a) Experimental and (b) theoretical dynamic response of device (D2) showing softening nonlinearity at low actuation drive and mixed nonlinear response (softening followed by hardening) at higher actuation drive.

The change in diaphragm stress due to annealing increases the fundamental vibrational mode's resonant frequency from 64.2kHz to 169.1kHz. A similar tuning in resonant frequency due to change in strain has been reported in doubly clamped membranes[44], beams and PMUTs[45,46]. The resonant mode at 169.1kHz is confirmed to be fundamental mode by mapping the mode shape (fig. S3). The dynamic response of D3 before and after annealing is shown in fig. 5(b) and (c) respectively. Before annealing, the static displacement was small and the cubic nonlinearity played a dominant role leading to the experimentally observed hardening nonlinear response. After annealing, the large static displacement gives rise to quadratic nonlinearities. For the measured displacements, effect of this quadratic nonlinearity dominates the effects of cubic nonlinearity and a softening nonlinear response is observed. The experimentally observed response is consistent with prediction of the numerical simulations. Unlike the dynamic response of device D2, where mixed nonlinearity was observed, we only observe softening nonlinearity in device D3 after annealing. We believe this is the result of large deformation observed in device D3 after annealing. The larger deformation in D3, compared to D2, gives rise to a larger contribution to quadratic nonlinearity, leading to softening nonlinearities even at higher actuation drives. It is likely that device D3 would also exhibit mixed nonlinearities at higher displacement amplitudes. These higher actuation voltages were close to the device damage threshold and thus were not viable.

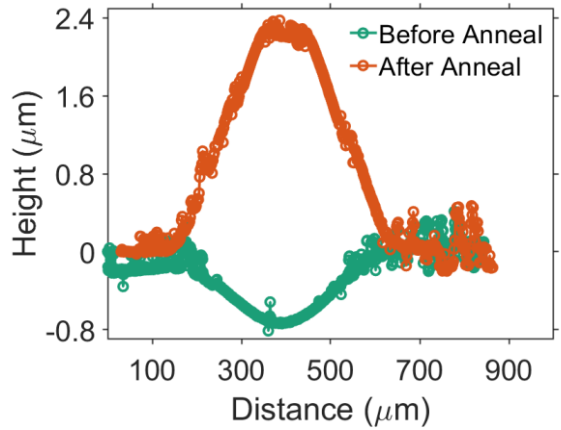

(a)

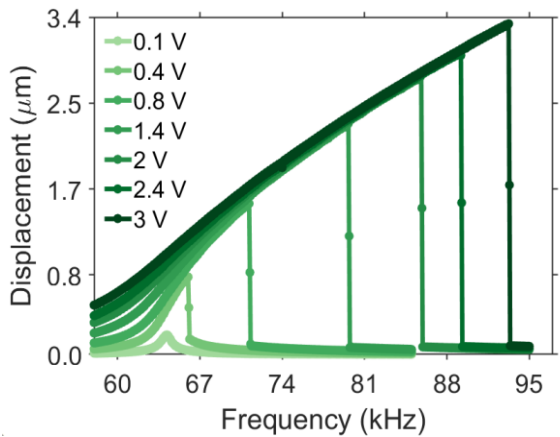

(b)

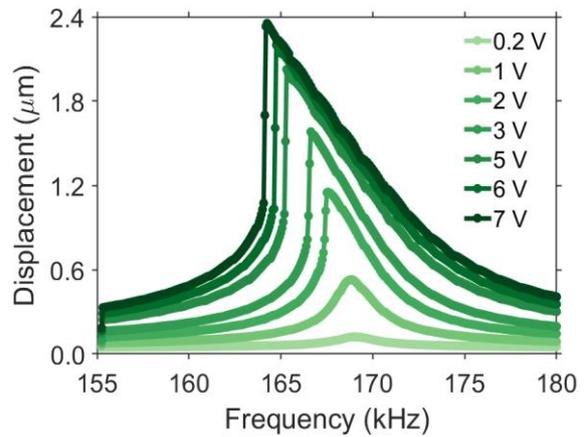

(c)

**Figure 5**: (a) Initial static displacement of the device (D3) before (green) and after (orange) annealing. Annealing leads to an increase in the static displacement of the diaphragm due to a change in the membrane stress. (b) Experimental frequency response curve of the fundamental mode before annealing demonstrating hardening nonlinearity with increasing actuation drive. (c) Experimental frequency response of the fundamental mode of the same device after annealing shows softening nonlinearity.

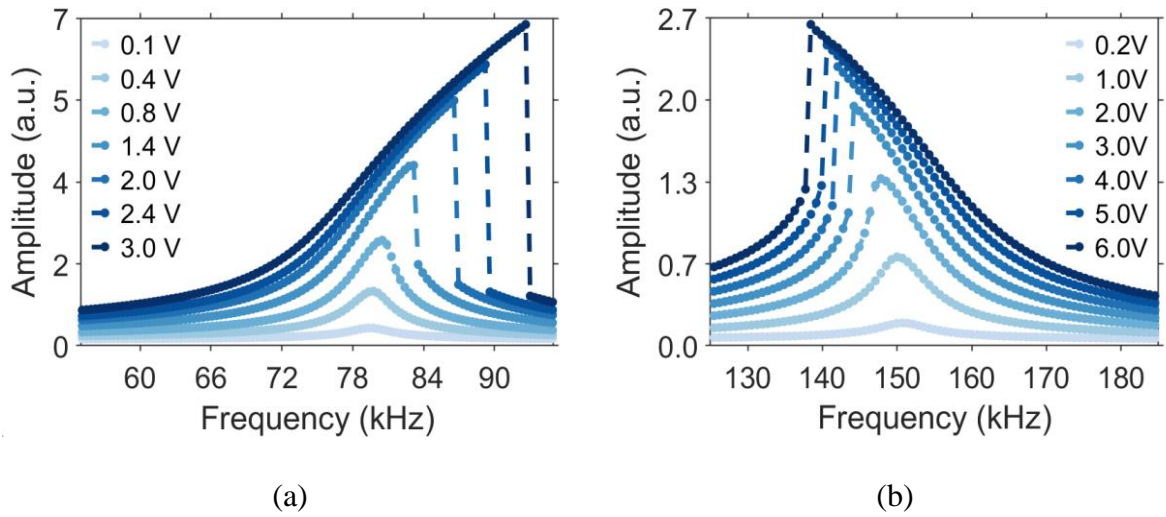

(a) (b)

**Figure 6**: (a) Numerical simulation of device (D3) with increasing actuation voltage showing (a) hardening nonlinearity before annealing with static displacement 0.55μm (0.11% of diameter) and (b) softening nonlinearity after annealing with static displacement 2.23μm (0.45% of diameter).

The above results clearly illustrate the effect of the initial static displacement of the device on its dynamics. To highlight this, fig. 7a shows the numerically simulated frequency response at constant voltage $(V_{ac} = 6\,V)$ for three different static displacements. For small static displacement (0.55μm (0.11% of diameter)), the effect of deformation on the dynamics is small and the device shows the hardening behaviour. For intermediate range of static displacement between 0.85μm (0.17% of diameter) to 1.2μm (0.24% of diameter), a mixed nonlinear response is observed, indicating contributions from both deformation and midplane stretching. For large static displacement of 1.5μm (0.30% of diameter) and above, only softening nonlinear response is observed. The transition from hardening nonlinearity to mixed nonlinearity and then to softening nonlinearity depends on the device parameters. Thus, changing the static displacement can significantly alter the dynamic response of the device. Figure 7b shows the ratio of the theoretically extracted quadratic to cubic nonlinearity coefficients with the initial static displacement of the PMUT. Here, the ratio increases with the static displacement. This leads to a change in the dynamic response of the PMUT from hardening to mixed and ultimately to softening nonlinearity.

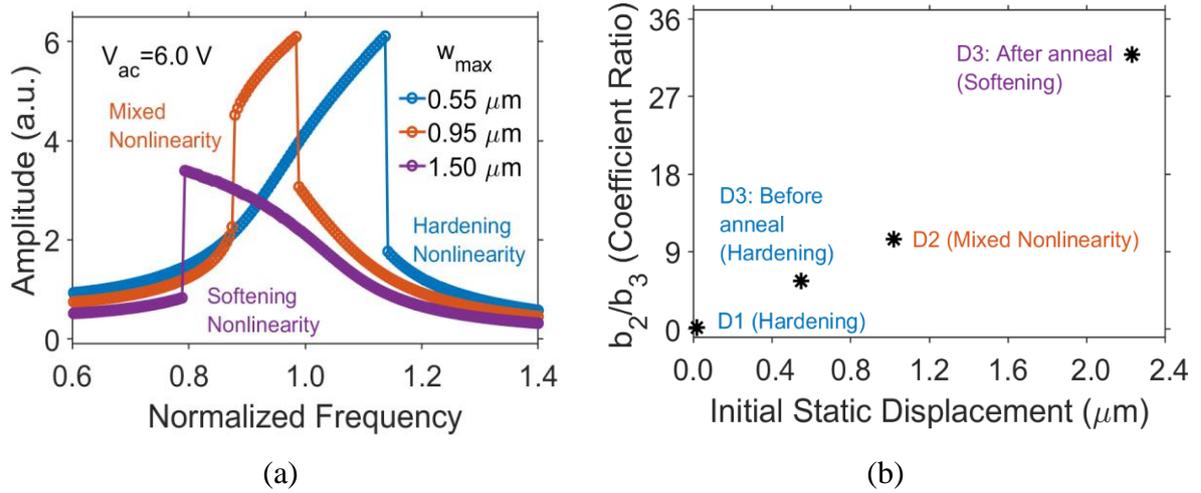

(a) (b)

**Figure 7**: (a) Simulated response of device showing hardening, mixed and softening nonlinearity for different initial static displacement. (b) The variation of ratio of coefficients of quadratic and cubic nonlinearity with the static displacement of the PMUT.

## Conclusions

In conclusion, we demonstrate the effect of initial curvature of ZnO PMUTs on their dynamic response. We find that the ZnO PMUTs with nominally flat diaphragms show the hardening nonlinearity when actuated with sufficiently large voltages. For the intermediate range of the static displacement, devices exhibit the mixed nonlinear behaviour and for large static displacement, devices exhibit only softening behaviour. We describe a theoretical model to understand the response of these devices and verify the experimental results through numerical simulations. The results indicate that the static displacement can be used to tune the quadratic and the effective nonlinearity of the device, thus altering its mechanical response. A careful future study on the effect of temperature on the diaphragm stress gradient leading to a change in the device's static displacement could enable the engineering of nonlinearities in PMUTs.

# Acknowledgement

N.A. acknowledges fellowship support under Visvesvaraya Ph.D. Scheme, Ministry of Electronics and Information Technology (MeitY), India. P.S. acknowledges Indian Institute of Science, Bangalore for funding through IOE scheme initiated by Government of India. We acknowledge funding support from MHRD, MeitY and DST Nano Mission through NNetRA. We thank Dr. Chandan Samanta and Dr. Sudhanshu Tiwari for helpful discussions.

[#]N.A. and P.S. contributed equally to this work.

# References

1. Naseri, H. & Homaeinezhad, M. R. Improving measurement quality of a MEMS-based gyro-free inertial navigation system. *Sensors Actuators A Phys.* **207**, 10–19 (2014).

2. Eom, C.-B. & Trolier-McKinstry, S. Thin-film piezoelectric MEMS. *Mrs Bull.* **37**, 1007–1017 (2012).

3. Wang, L. *et al.* Rapid and ultrasensitive electromechanical detection of ions, biomolecules and SARS-CoV-2 RNA in unamplified samples. *Nat. Biomed. Eng.* **6**, 276–285 (2022).

4. Korvink, J. G. *MEMS : A Practical Guide to Design , Analysis and Applications*. (2010).

5. Przybyla, R. J. *et al.* In-air rangefinding with an aln piezoelectric micromachined ultrasound transducer. *IEEE Sens. J.* **11**, 2690–2697 (2011).

6. Xu, R., Zhou, S. & Li, W. J. MEMS Accelerometer Based Nonspecific-User Hand Gesture Recognition. *IEEE Sens. J.* **12**, 1166–1173 (2012).

7. Siu, C.-P.-B., Zeng, H. & Chiao, M. Magnetically actuated MEMS microlens scanner for in vivo medical imaging†. *Opt. Express* **15**, 11154–11166 (2007).

8. Dangi, A. *et al.* A Photoacoustic Imaging Device Using Piezoelectric Micromachined Ultrasound Transducers (PMUTs). *IEEE Trans. Ultrason. Ferroelectr. Freq. Control* **67**, 801–809 (2020).

9. Gupta, H., Nayak, B., Ashok, A. & Pratap, R. Data-Over-Sound With PMUTs. *IEEE Open J. Ultrason. Ferroelectr. Freq. Control* **2**, 152–161 (2022).

10. Xie, J., Lee, C. & Feng, H. Design, Fabrication, and Characterization of CMOS MEMS-Based Thermoelectric Power Generators. *J. Microelectromechanical Syst.* **19**, 317–324 (2010).

11. Suzuki, K., Higuchi, K. & Tanigawa, H. A silicon electrostatic ultrasonic transducer. *IEEE Trans. Ultrason. Ferroelectr. Freq. Control* **36**, 620–627 (1989).

12. Park, K. K., Oralkan, O. & Khuri-Yakub, B. T. A comparison between conventional and collapse-mode capacitive micromachined ultrasonic transducers in 10-MHz 1-D arrays. *IEEE Trans. Ultrason. Ferroelectr. Freq. Control* **60**, 1245–1255 (2013).

13. Wang, M. & Zhou, Y. Design of piezoelectric micromachined ultrasonic transducers (pMUTs) for high pressure output. *Microsyst. Technol.* **23**, 1761–1766 (2017).

14. Sammoura, F., Smyth, K. & Kim, S.-G. Optimizing the electrode size of circular bimorph plates with different boundary conditions for maximum deflection of piezoelectric micromachined ultrasonic transducers. *Ultrasonics* **53**, 328–334 (2013).

15. Dausch, D. E. *et al.* In vivo real-time 3-D intracardiac echo using PMUT arrays. *IEEE Trans. Ultrason. Ferroelectr. Freq. Control* **61**, 1754–1764 (2014).

16. Lu, Y. *et al.* Ultrasonic fingerprint sensor using a piezoelectric micromachined ultrasonic transducer array integrated with complementary metal oxide semiconductor electronics. *Appl. Phys. Lett.* **106**, 263503 (2015).

17. Jiang, X. *et al.* Ultrasonic Fingerprint Sensor With Transmit Beamforming Based on a PMUT Array Bonded to CMOS Circuitry. *IEEE Trans. Ultrason. Ferroelectr. Freq. Control* **64**, 1401–1408 (2017).

18. Ardito, R., Bertarelli, E., Corigliano, A. & Gafforelli, G. On the application of piezolaminated composites to diaphragm micropumps. *Compos. Struct.* **99**, 231–240 (2013).

19. Farshchi Yazdi, S. A., Corigliano, A. & Ardito, R. 3-D Design and Simulation of a Piezoelectric Micropump. *Micromachines* **10**, (2019).

20. Smyth, K., Sodini, C. & Kim, S.-G. High electromechanical coupling piezoelectric micro-machined ultrasonic transducer (PMUT) elements for medical imaging. *2017 19th Int. Conf. Solid-State Sensors, Actuators Microsystems* 966–969 (2017).

21. Chen, X., Xu, J., Chen, H., Ding, H. & Xie, J. High-Accuracy Ultrasonic Rangefinders via pMUTs Arrays Using Multi-Frequency Continuous Waves. *J. Microelectromechanical Syst.* **28**, 634–642 (2019).

22. Chen, X. *et al.* A High-Accuracy in-Air Reflective Rangefinder Via PMUTS Arrays Using Multi-Frequency Continuous Waves. in *2019 20th International Conference on Solid-State Sensors, Actuators and Microsystems & Eurosensors XXXIII (TRANSDUCERS & EUROSENSORS XXXIII)* 154–157 (2019). doi:10.1109/TRANSDUCERS.2019.8808651

23. Nazemi, H. *et al.* Mass Sensors Based on Capacitive and Piezoelectric Micromachined Ultrasonic Transducers—CMUT and PMUT. *Sensors* **20**, (2020).

24. Roy, K. *et al.* Fluid Density Sensing Using Piezoelectric Micromachined Ultrasound Transducers. *IEEE Sens. J.* **20**, 6802–6809 (2020).

25. Choi, S., Kim, S., Yoon, Y. & Allen, M. G. Exploitation of Nonlinear Effects for Enhancement of the Sensing Performance of Resonant Sensors. in *TRANSDUCERS 2007 - 2007 International Solid-State Sensors, Actuators and Microsystems Conference* 1745–1748 (2007). doi:10.1109/SENSOR.2007.4300490

26. Kacem, N., Arcamone, J., Perez-Murano, F. & Hentz, S. Dynamic range enhancement of nonlinear nanomechanical resonant cantilevers for highly sensitive NEMS gas/mass sensor applications. *J. Micromechanics Microengineering* **20**, 45023 (2010).

27. Samanta, C., Arora, N. & Naik, A. K. Tuning of geometric nonlinearity in ultrathin nanoelectromechanical systems. *Appl. Phys. Lett.* **113**, 113101 (2018).

28. Yuksel, M. *et al.* Nonlinear Nanomechanical Mass Spectrometry at the Single-Nanoparticle Level. *Nano Lett.* **19**, 3583–3589 (2019).

29. Fetisov, Y. K., Burdin, D. A., Chashin, D. V & Ekonomov, N. A. High-Sensitivity Wideband Magnetic Field Sensor Using Nonlinear Resonance Magnetoelectric Effect. *IEEE Sens. J.* **14**, 2252–2256 (2014).

30. Akhbari, S. *et al.* Highly responsive curved aluminum nitride pMUT. in *2014 IEEE 27th International Conference on Micro Electro Mechanical Systems (MEMS)* 124–127 (2014). doi:10.1109/MEMSYS.2014.6765589

31. Cho, H., Bergman, L. A., Yu, M. F. & Vakakis, A. F. Intentional nonlinearity for

design of micro/nanomechanical resonators. in *Nanocantilever Beams: Modeling, Fabrication and Applications* 137–192 (Pan Stanford Publishing Pte. Ltd., 2016).

32. Mahmoodi, S. N., Jalili, N. & Daqaq, M. F. Modeling, Nonlinear Dynamics, and Identification of a Piezoelectrically Actuated Microcantilever Sensor. *IEEE/ASME Trans. Mechatronics* **13**, 58–65 (2008).

33. Gusso, A. Nonlinear damping in doubly clamped beam resonators due to the attachment loss induced by the geometric nonlinearity. *J. Sound Vib.* **372**, 255–265 (2016).

34. Singh, P. & Yadava, R. D. S. Effect of viscous axial loading on vibrating microcantilever sensors. *J. Phys. D. Appl. Phys.* **52**, 345301 (2019).

35. Amabili, M. Nonlinear vibrations of doubly curved shallow shells. *Int. J. Non. Linear. Mech.* **40**, 683–710 (2005).

36. Jallouli, A., Kacem, N. & Lardies, J. *Nonlinear static and dynamic behavior of an imperfect circular microplate under electrostatic actuation. Proceedings of the ASME Design Engineering Technical Conference* **4**, (2017).

37. Jallouli, A., Kacem, N. & Lardies, J. Investigations of the effects of geometric imperfections on the nonlinear static and dynamic behavior of capacitive micomachined ultrasonic transducers. *Micromachines* **9**, (2018).

38. Kumar, R., Tiwari, S., Thakur, V. & Pratap, R. Growth of ultrafast, super dense ZnO nanorods using microwaves for piezoelectric MEMS applications. *Mater. Chem. Phys.* **255**, 123607 (2020).

39. Kumar, R., Tiwari, S. & Pratap, R. Significant Enhancement in Operational Bandwidth of ZnO PMUTs due to the Simultaneous Existence of Softening and Hardening Nonlinearity. in *2020 IEEE International Ultrasonics Symposium (IUS)* 1–4 (2020). doi:10.1109/IUS46767.2020.9251421

40. Smyth, K. & Kim, S.-G. G. Experiment and simulation validated analytical equivalent circuit model for piezoelectric micromachined ultrasonic transducers. *IEEE Trans. Ultrason. Ferroelectr. Freq. Control* **62**, 744–765 (2015).

41. Bose, T. & Mohanty, A. R. Large amplitude axisymmetric vibration of a circular plate having a circumferential crack. *Int. J. Mech. Sci.* **124–125**, 194–202 (2017).

42. Timoshenko, S. & Timošenko, S. P. *Theory of Plates and Shells*. (McGraw-Hill Book Company, Incorporated, 1940).

43. Dangi, A. & Pratap, R. System level modeling and design maps of PMUTs with residual stresses. *Sensors Actuators A Phys.* **262**, 18–28 (2017).

44. Samanta, C., Arora, N., V., K. K., Raghavan, S. & Naik, A. K. The effect of strain on effective Duffing nonlinearity in the CVD-MoS2 resonator. *Nanoscale* **11**, 8394–8401 (2019).

45. Sadeghpour, S., Kraft, M. & Puers, R. Design and fabrication strategy for an efficient lead zirconate titanate based piezoelectric micromachined ultrasound transducer. *J. Micromechanics Microengineering* **29**, (2019).

46. Ross, G. *et al.* The impact of residual stress on resonating piezoelectric devices. *Mater. Des.* **196**, (2020).